# 100 GHz Micrometer-compact broadband Monolithic ITO Mach–Zehnder Interferometer Modulator enabling 3500 times higher Packing Density

Yaliang Gui[1], Behrouz Movahhed Nouri[1], Mario Miscuglio[1], Rubab Amin[1], Hao Wang[1], Jacob B. Khurgin[2], Hamed Dalir[1], and Volker J. Sorger[1]
[1]George Washington University, 800 22nd Street NW, Washington, DC 20052, USA
[2]Johns Hopkins University, Baltimore MD 21208, USA
+Corresponding authors: hdalir@gwu.edu , sorger@gwu.edu

**Abstract:** Electro-optic modulators provide a key function in optical transceivers and increasingly in photonic programmable Application-Specific Integrated Circuits (ASICs) for machine learning and signal processing. However, both foundry-ready silicon-based modulators and conventional material-based devices utilizing Lithium-niobate fall short in simultaneously providing high chip packaging density and fast speed. Current-driven ITO-based modulators have the potential to achieve both enabled by efficient light-matter interactions. Here, we introduce micrometer-compact Mach Zehnder Interferometer (MZI)-based modulators capable of exceeding 100 GHz switching rates. Integrating ITO-thin films atop a photonic waveguide, one can achieve an efficient $V_\pi L$=0.1 V·mm, spectrally broadband, and compact MZI phase shifter. Remarkably, this allows integrating more than 3500 of these modulators within the same chip area as only one single silicon MZI modulator. The modulator design introduced here features a holistic photonic, electronic, and RF-based optimization and includes an asymmetric MZI tuning step to optimize the Extinction Ratio (ER)-to-Insertion Loss (IL) and dielectric thickness sweep to balance the trade-offs between ER and speed. Driven by CMOS compatible bias voltage levels, this device is the first to address next-generation modulator demands for processors of the machine intelligence revolution, in addition to the edge- and cloud computing demands as well as optical transceivers alike.

**Keywords:** electro-optic modulators, silicon photonics, optical, transparent conductive oxides, indium tin oxide, asymmetric power splitter, Mach-Zehnder interferometer

# 1 Introduction

Electro-optical modulators (EOMs) tune the light intensity or phase at certain wavelength. The Mach-Zehnder based optical modulator change the light intensity or phase by having the electro-optical effect on the active arms to shift the transmission spectral, results in an amplitude modulation or phase modulation signal. Currently for high-speed data transmission in integrated silicon photonics suffers from large footprint, high insertion loss, high energy consumption, low extinction ratio due to weak electro-refractivity. For example, a current available foundry doped silicon based MZI modulator (MZM), while being reasonably fast ($f_{3db}$~27 GHz), require a large footprint of 0.74 mm$^2$.[1] Using an electronic processor die as a footprint-guide (1 cm$^2$ = 100 mm$^2$), the number of Photonic Integrated Circuit (PIC) components on-chip is severely limited to about 100-1000 components as compared to electronic counterparts. The MZI insertion loss < 3dB, extinction ratio > 15 dB and $V_\pi$ L = 90 V·mm[2]. With looming technological advances such as the internet-of-things [3], network-edge computing [3, 4], neuromorphic computing [5, 6], chip-scale light detection and ranging (LiDAR) [7]; the need for increasingly efficient and miniaturized EOMs is relentless.

For example, the photonic processors scale differently than electronics; by a) multiplexing options and b) synergistic complexity scaling reductions to algorithms. The algorithm-to-hardware mapping, we can look at performing complex convolutions optically as much simpler dot-product multiplications in the Fourier domain. While this is known in signal processing, the required overhead-heavy Fourier Transportation (FT) is performed effortlessly (passively) by an optical lens (either in free space or on a PIC). Mathematical operations scale with a nominal dimensionality factor N, reflecting the runtime complexity (under brute-force, that is, meaning without parallelization strategies). Tensor operations and convolutions, for example, scale with N3 and N4, respectively, assuming a squared matrix of size N. Why is this relevant for modulators? Because the performance (typically stated in throughput per second per power, i.e., TOPS/W=TOP/J) of an accelerator is a direct function of the speed of each modulator and the number of modulators on the application specific integrated circuit (ASIC) PIC. Hence, being fast and compact becomes critically crucial for next-generation photonic ASICs.

To address these relevant information processing trends using reconfigurable photonic circuits and hence modulators, several EOM devices schemes have been proposed to enhance Si-based devices' performance only to reveal inadequate improvements beyond the basic and selectively doped junctions[8]. In this view, alternative electro-optic materials for active modulation are sought, offering higher index tunability, seamless integration on the Si platform, ease of relevant processing, and fast modulation in a compact form factor. A plethora of materials have been investigated for this purpose in recent years ranging from the traditional III-V materials [9], polymers [10], ferroelectric materials [11] to more exotic 2-dimensional materials, including graphene [12], transition metal dichalcogenides[13], etc. In this regard, a class of materials that have received growing interest is transparent conducting oxides (TCOs). TCOs usually feature a fairly wide-bandgap with degenerate doping levels to facilitate electrical conduction while remaining transparent in the widely utilized telecom C-band.

Indium tin oxide (ITO) is the most widely used TCO material in industry and research alike; it offers current-based modulation in an electrical capacitive arrangement based on similar carrier dispersion dynamics as Si. Compared to other materials, integrating ITO in chip-scale modulation schemes presents the following advantages: (i) Availability of at least 3-4 orders of magnitude higher carrier density compared to Silicon; (ii) Dramatic effect of change in carrier concentration on its optical index, which arises from the small permittivity of the material, i.e., $\partial n = \partial \epsilon^{1/2} \sim \partial \epsilon / 2\epsilon^{1/2}$; (iii) The presence of an epsilon-near-zero (ENZ) region in the tolerable carrier concentration range within the electrostatic gating constraints; and (iv) Potential CMOS compatibility, which renders it integrable in the mature Si process flow in a monolithic manner [6], [14]–[17]. The latter is a critical factor, enabling mass scale integration and moving the modulators out of the lab into technology development and production.

Pioneering works in ITO-based absorption modulation influenced interest in its efficacy as a phase modulator stemming from the well-known Kramers-Kronig (K-K) relations [18]. Recently, we identified that phase tuning efficacy surpasses that of absorption tuning for ITO in associated performance metrics [19]. Indeed, ITO is an exciting alternative for highly performing phase modulation as such has been manifested by our recent works supporting half-wave voltage and active device length products, $V_\pi L$ down to 63 V·μm for a lateral configuration device [20]. We have also recently demonstrated noteworthy results in terms of $V_\pi L$ in traditional photonic and plasmonic designs obtaining 520 V·μm and 95 V·μm, respectively [17], [21], [22]. A comparison of disparate performance metrics for different material and model-based schemes found in recent literature is presented for reference (Table 1). Different phase modulation schemes have been investigated with ITO thus far to include traveling wave-based solutions as well as cavity-based approaches[13–15], [17–24]. In our comparison, cavity-based approaches are not considered since mirroring resonators can offer a higher ER with a slight change in the index but come at the cost of significantly reducing the spectral bandwidth owing to the finesse of the cavity. Additionally, the feedback schemes (e.g., ring circumference or photonic crystal mirror sections) cannot be excluded from the footprint calculations of the active device as these specify the needed finesse of the cavity, which is integral to the device performance of said devices integrated within them.

On the other hand, traveling wave-based schemes such as MZIs are single-pass arrangements that can deliver the full broadband optical response the chosen optical mode offers depending on the arms' imbalance. Furthermore, the entire MZI area can be reduced to only the active device footprint plus the spacing for another MZI arm and Y-couplers without any conceivable effect to hinder the device's performance. This can, de facto, be achieved in a minuscule footprint owing to lithographic advancement in recent years paired with cutting-edge fabrication facilities.

Recapitulating the efforts to minimize the real estate on the chip, plasmonics can serve as a reasonable alternative approach to traditional photonic packaging. Plasmonics effectively squeezes the light in the optical mode to enhance Light-Matter Interaction (LMI) with the active modulation material realizing ultrashort devices with improved modulation metrics. Such enhancements in active performance do inherently accompany additional loss in the form of IL due to the high modal mismatch when coupling back and forth from the underlying Si waveguide. These losses can be partially mitigated by employing modal transitional architecture within the coupling regions in the vicinity of the active device, for example, tapered waveguide sections for plasmonic slot designs. We have shown that the enhanced LMI alleviates the high loss per unit length encountered in plasmonic, and the performance benefits outweigh the initial surcharge of IL if the latter is managed to retain at a tolerable limit [17, 24]. The high real index tuning in ITO accompanies a considerable modulating loss directly resulting from the imaginary part of the index via K-K relations, and this effect is even stronger in the plasmonic mode-based setups as the confinement can reach at least one order of magnitude more than the photonic case.

Simple integration of an ITO-based modulator in one arm of an MZI thus limits attainable extinction and refrains from a linear operation, inducing substantial chirp accrual in the process. We had mitigated this issue by imposing some static loss in the form of a passive metallic contact on the other unmodulated arm of the MZI in our previous works with a view to minimizing the loss imbalance in both arms [15, 19, 20]. The static passive metallic method balances the output by introducing loss on the passive arm, increasing IL. One alternative approach is to design the input Y-junction to deliver more power to the active arm to minimize the imbalance resulting upon active switching, decrease IL, and reduce the number of required fabrication steps. What's more, this asymmetric Y-junction design can also be used to achieve low IL by delivering less power to the active arm. Whether to provide more power to the active arm to minimize the imbalance or deliver less power to increase the imbalance but decrease the IL depends on the applications. (Details in Supporting Information (SI))

In this work, we propose an ITO-plasmon-based asymmetric MZI modulator to solve the aforementioned problems. The numerical simulation results show the modulator affords the advantages of low bias voltage and compact device size, low IL, adequate ER, low electrical resistance leading to multi-GHz operational speeds, etc.

Our design strategy, which can be applied to other material frameworks, enables unitary ER/IL without sacrificing footprint (<5µm phase shifter, and $2 \cdot 10^{-4}$ mm$^2$). Additionally, the crafted material properties and appropriately tuned thicknesses lead to superior electrostatic performance, aiding unprecedented switching speed (up to 100 GHz) and one order of magnitude lower low energy consumption (380 fJ/bit) with respect to state of the art (Table 1). Such performance heralds a new device makes it a strong candidate to become the fundamental building block of the next-generation photonic platform with immediate applications in neuromorphic computing [25] and programmable photonic circuits [26].

**Table 1**: Figure of merit comparison for different type Mach-Zehnder interferometric modulators

| Material | Mechanism | Length ($\mu m$) | ER (dB) | IL (dB) | Speed (GHz) | $V_\pi L$ ($V \cdot \mu m$) | Energy consumption (fJ/bit) | Ref. |
|---|---|---|---|---|---|---|---|---|
| LN | Pockels effect | 20000 | 30 | 0.4 | 45 | 2800 | 0.37 | [27] |
| Si | plasma dispersion effect | 15000 | 3.8 | 10 | 10 | 3300 | 138000 | [28] |
| Polymer | Pockels effect | 1000 | 10 | 6 | 12.5 | 500 | 0.7 | [29] |
| ITO | plasma dispersion effect | 1.4 | 8 | 6.7 | 1.1 | 95 | 2100 | [15] |
| ITO | plasma dispersion effect | 4.7 | 3* | 2.9* | 100* | 108 | 380* | This Work |

**Note** numbers with '*' its calculated value.

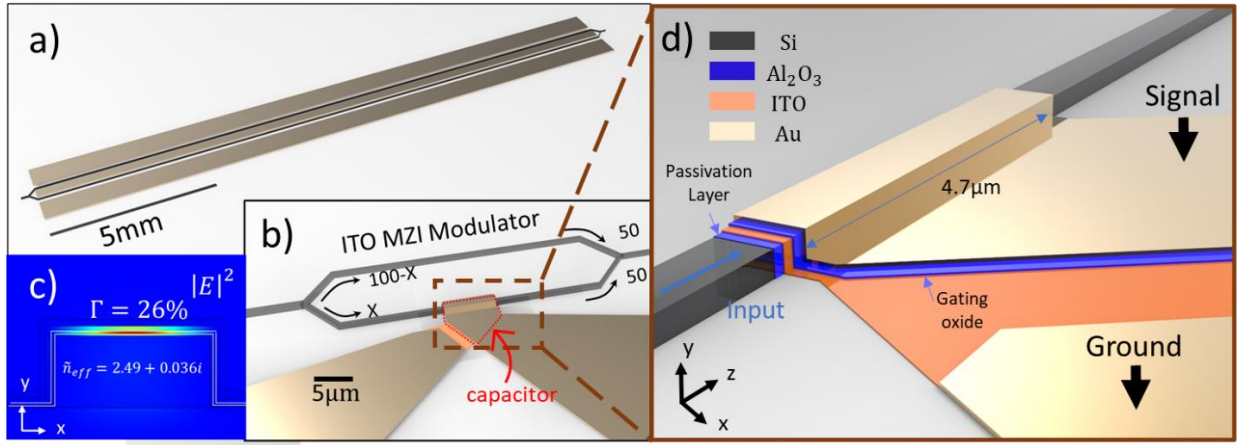

**Fig. 1:** ITO-based plasmonic MZI modulator on Si photonic platform. (a) Footprint comparison between ITO MZI modulator and traditional MZI modulator. The ITO-based plasmonic MZI modulator with input and output grating couplers is 1mm long and 165µm in width, including the contact pads. The size of ITO-based plasmonic MZI is 33µm in length and 6µm in width. The distance between the grating coupler and the device is longer than 500µm to facilitate measurement. (b) Perspective view of the Mach-Zehnder structure with the active biasing contacts. The power splitting ratio for the active arm is X, and the splitting ratio for the passive arm is 100-X. (c) Electric field distribution by performing FEM at 1550nm under +3.5V in the cross-sectional structure for a z cutline along the central region of the Si waveguide (width: 500nm; height: 220nm). (d) Composition layers of ITO MZI modulator. ($T_{passivation}$ = 5nm; $T_{ITO}$ = 10nm; $T_{gating}$ = 15nm; $T_{Au}$ = 30nm)

## 2 Results and discussion

Modulators are inherently optimization-sensitive devices. There is no single best design as many parameters are cross- and inter-dependent (Fig. 3a). For example, thick gating oxide can reduce the capacitance and result in higher speed; however, it undermines the confinement factor, which leads to weak modulation depth. As such, we bound ourselves by a set of performance requirements. While these are arbitrarily chosen, they are informed by the rationale above for next-generation modulators with an option for ASPICs and defined as follows: ER > 3 dB, IL < 3 dB, (i.e., ER/IL > 1), 3dB >100GHz, footprint > $10^{-3}$ mm2, and energy consumption ~300 fJ/bit. The following discussions revolve around this goal.

First, we introduce the scheme of the phase shifter and carefully analyze material properties under different voltages. Thus, the phase shift and electrical performance, such as system resistance, capacitance, and speed, are obtained for devices with different lengths and gating oxide thickness. As discussed above, gating oxide thickness is critical for speed and modulation depth. However, different Secondly, we discuss the above-introduced asymmetric splitting ratio of the front-end Y-junction of the MZI as a secondary application wised optimization strategy to reach an optimized point in terms of trade-offs between ER and IL within the constraints of small phase shifts.

A 5 nm thin film of $Al_2O_3$ is placed on top of the SOI MZI as a passivation layer, enhancing the grating coupler's coupling efficiency. (Fig. 1d) . Light modulation is achieved by phase-shifting the mode in one arm of the MZI modulator via electrostatically tuning the free carriers of a 10 nm ITO layer in a parallel plate capacitor configuration. Additionally, the metal layer used as the top plate of the capacitor generates a plasmonic mode, enhancing light-matter interaction and obtaining strong, effective modal index variation upon carrier concentration tuning in the ITO film (Fig. 1c). The thickness of the passivation layer and ITO thin film is chosen based on the fabrication techniques and experimental experience to get a closer approximation to the real situation. Although thinner passivation layer results in a higher confinement factor, passivation layer with thickness lower than 5nm has issues such as unevenness, discontinuity, and difficulty in measurement. The same with ITO thin films. With the improvement of the fabrication technique, high-quality ultra-thin dielectric and ITO films can be used to increase the confinement factor and shorten the device length. As further discussed, the gating oxide thickness is selected to compromise speed performance while having 3dB ER. It's worth noting that although dielectrics with a high dielectric constant can stand for high voltage, its high dielectric constant can also undermine the speed by increasing the capacitance. (SI)

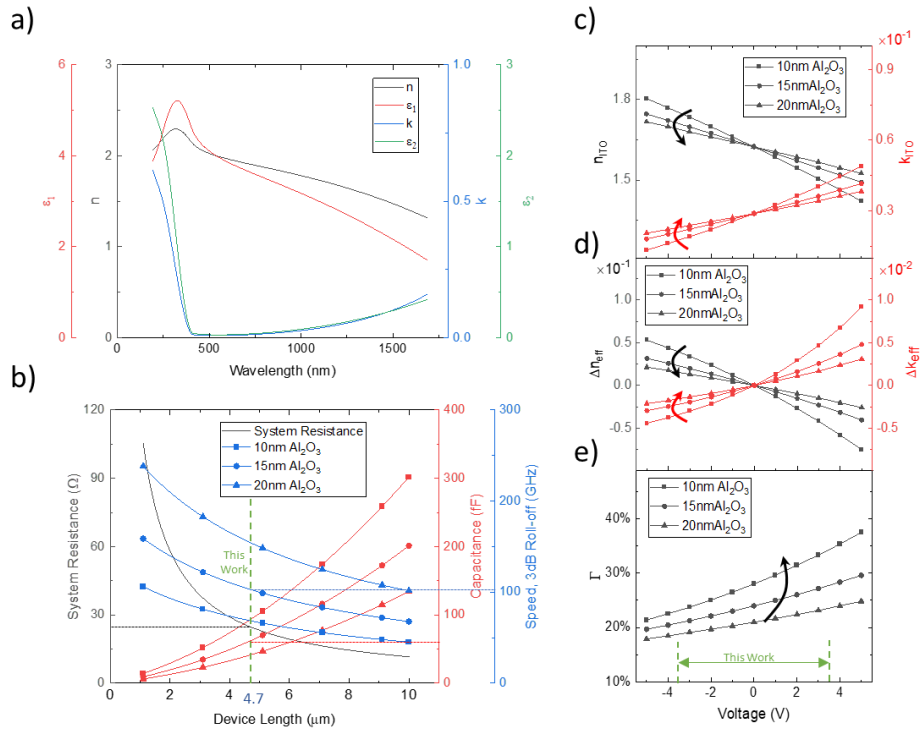

**Fig. 2:** Tailored material and design parameters of the phase shifter. a) Dispersive spectrum of 10nm ITO thin film measured by ellipsometry b) Resistance and speed of ITO MZI modulator with a gating oxide thickness of 10nm, 15nm, and 20nm. c) Refractive indices and extinction coefficients of 10nm in thick and 4.7um in lengthITO thin film for various dielectric thicknesses under different voltages at 1550nm. d) Effective mode index differences change with dielectric thickness under varied voltages at1550nm. e) Confinement factor Γ varies with dielectric thickness and voltages at1550nm.

The modulation mechanism in ITO is free-carrier dispersion-based accumulation/depletion. The optical properties can be controlled by well-timed adjusted deposition techniques to concurrently limit off-state optical losses and keep on-state away from epsilon-near-zero higher losses. (Fig.2a) [16] The Drude oscillator is used to describe the free-carrier absorption, while the Tauc-Lorentz equations are used to describe the effects of electronic transitions in the short wavelength.[30] In this work, refractive indexes are calculated from carrier concentration using Drude model. The refractive index of the unbiased ITO film is obtained from the reported carrier concentration, $2.07 \times 10^{20} cm^{-3}$. [15] The depleted or accumulated carriers and corresponding refractive indexes in the active ITO region upon capacitive operation under the bias using Thomas–Fermi screening model. [32] The real part of refractive index the decreases for increasing positive voltages corresponding to a blue shift in wavelength. In contrast, the imaginary part increases towards a more metalic characteristic (Fig.2c). Assuming a fixed, CMOS compatible voltage of ±3.5V, carrier dynamic charge accumulation $\Delta Q$ is inversely proportional to the dielectric thickness. ( $\Delta Q = C \times \Delta V$, and $C = \epsilon_0 \epsilon_r A/d$ , where $\Delta Q, C, \Delta V, \varepsilon_r, A, d$ denoting the accumulated charge, capacitance, changed voltage, the relative dielectric constant, area of two overlap plates, dielectric thickness). In other words, under the same voltage bias, a thinner dielectric can influence an extensive refractive index and extinction coefficient variation.

To optimize the modulation performance of the proposed phase shifter, compensate for losses, and tune the design parameters, we study the effective mode index variation by performing eigenmode simulation using FEM analysis. The variation of the effective refractive $\Delta n_{eff}$ is a monotonically decreasing function of the applied voltage (Fig.2d). The slopes of the curves are governed by the oxide thickness, which is responsible for modulation depth per unit length and loss per unit length. With a superliner proportionality, the electric field distribution is associated with the propagating mode, thus altering the overlap factor within the active layer (Fig.2e). A thinner dielectric layer leads to a higher confinement factor, therefore, stronger carrier interaction and a higher ER. Under the accumulation state, $\Delta n_{eff}$ of device with 10nm gating oxide is two times larger than that of a device with 15nm gating oxide because of stronger light-matter interaction. However, it also induces higher loss, according to Kramers-Kronig relations. Furthermore, the working voltage is constrained by the breakdown voltage, and blindly pursuing high energy confinement is inadvisable. Once the phase shift and related absorption coefficient are obtained, we can define the phase shifter footprint according to the desired modulation depth and derive the modulation speed as a function of the oxide thickness. This is further discussed in figure 4.

The electrical properties of the phase shifter device are calculated from reported ITO resistivity of $2.9 \times 10^{-4} \Omega \cdot cm$ after thermal annealing. [28] The system resistance is obtained from COMSOL Multiphysics. (Fig. 2b) It decreases nonlinearly with the device length because ITO thin film is laid on the protruding silicon waveguide and the shape of the thin film is abnormal. Compared with the methods that calculate the ITO thin films as rectangular, calculating the result using the simulation is more convincing and reliable. The capacitance is calculated based on the shape that is closer to the real situation. And the nonlinearly decreased system resistances result in dispersive speeds.

Footprint has a strong impact on RC delay. Even if a larger overlap area of the two plates would considerably reduce the voltage to be applied or eventually improve modulation depth, providing π-shift for the same voltage applied, it would significantly compromise device speed. In contrast, thicker oxide, which can improve speed performance (lowering capacitance), would reduce trade-offs in terms of optical losses.

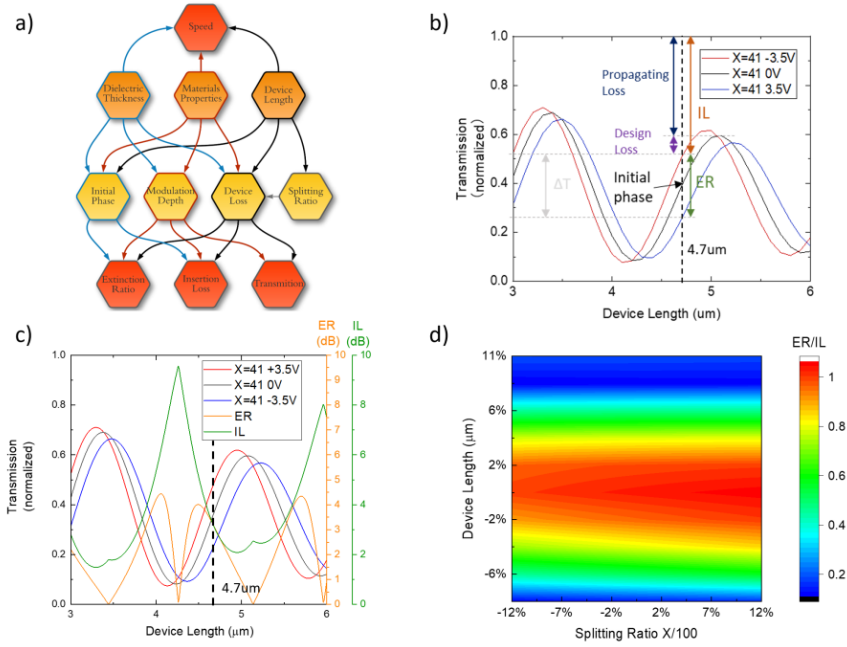

**Fig. 3:** a) Flow chart of Mach-Zehnder modulator analysis. Variables are dielectric thickness (gating oxide thickness), material properties, device length, and splitting ratio. The performances are ER, IL, transmission, and speed. b) Transmission of ON state, 0 bias state, and OFF state oscillate within the device with a dielectric thickness of 15nm under voltage of +3.5V, 0V, and -3.5V when splitting ratio is 41:59. Propagating loss is caused by absorption, while the design loss is caused by insufficient phase shift. c) Transmission, corresponding ER, and IL varies with device length. ER=3dB and IL =2.9dB when device length is 4.7um. d) Fabrication tolerance of device length and splitting ratio for the proposed device.

Here we distinguish amongst, propagating loss, defined by $\alpha = \frac{4\pi k_{eff}}{\lambda} = 0.27 dB/\mu m$ and design loss, caused by insufficient modulation depth (phase shift of $0.33\pi$). Ideally, for a complete $\pi$ shift device, design loss would be 0. The initial phase is dependent on the device length and the difference between effective mode indexes of two arms. (Fig.3b) In this work, we opted for a total phase shift of $0.33\pi$ ( $\Delta n_{eff} = 0.1516$ ), corresponding to 3dB modulation depth and total IL (sum of propagating loss and design loss) of 2.9 dB for an active region of just 4.7μm (Fig.3c).

Additionally, we verified the theoretical impact of fabrication tolerances on the phase shifter performance; with a $\pm 4\%$ tolerance in device footprint and over $\pm 12\%$ tolerance in splitting ratio (splitting ratio of $\pm 8\%$ is the current foundry tolerance), we can still achieve an ER/IL=0.8, demonstrating reasonably achievable robustness, certainly given foundry precision. (Fig.3d) To be noticed, the robustness is at the expense of unity decreasing or increasing of the ER and IL. The details about the ER and IL will be discussed in the following context.

Considering the constraints, we select a 15nm gating oxide which allows altering the ITO carrier concertation from $1.6 \times 10^{20} cm^{-3}$ to $2.5 \times 10^{20} cm^{-3}$ under ±3.5V electrostatic gating while still ensuring fast modulation (up to 100 GHz) and optimized losses against modulation (ER/IL=3/2.9=1.03). The −3dB roll-off is estimated through RC delay, given the capacitance of 62 fF (red dash line in Fig 2b) and the total resistance of 25 Ω, while dynamic switching energy is 380pJ.

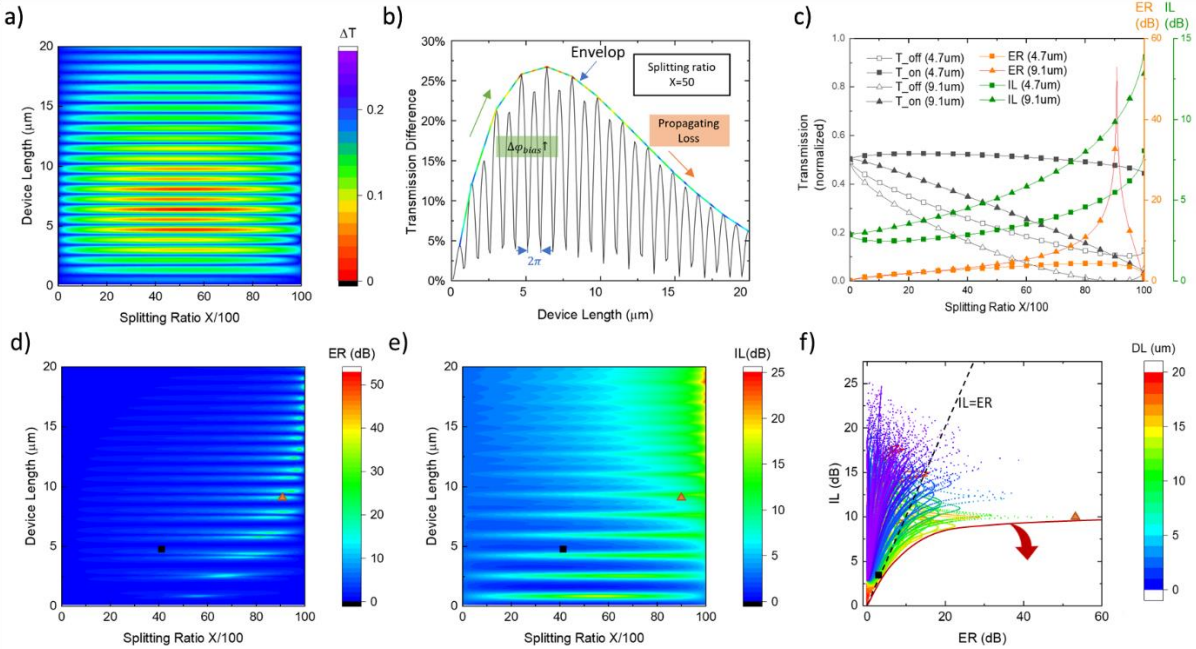

**Fig. 4:** The performance map of asymmetric plasmonic mode ITO-based MZI modulator with 15nm $Al_2O_3$ under ON (-3.5V) and OFF states (+3.5V). Squares represent the device with 3dB ER, 2.9dB IL, 25%ΔT, and a speed of 102GHz. Triangle represents the device with 53dB ER, 10dB IL, 10%ΔT, and speed of 71GHz. A) Transmission difference varies with splitting ratio X and device length 0 $\mu m$ to 20$\mu m$ (More details for photonic mode device is included in SI) Plasmonic designs with transmission differences higher than 25% locate around 5$\mu m$ because of high propagating loss of the longer device. B) Transmission difference of plasmonic mode varies with device length at splitting ratio 50:50. The transmission oscillates because of phase shift by both device length and bias. And the colorful envelope is manually added to show the transmission difference changes with the phase shift caused by the bias and propagating loss. C) Examples: transmissions under ON and OFF state, ER, and IL for devices with lengths of 4.7$\mu m$ (X=41) and 9.1$\mu m$ (X=91), respectively. D) ER of plasmonic mode and photonic device varies with the splitting ratio for different lengths. E) IL of plasmonic mode and photonic mode varies with the splitting ratio for different lengths. F) Performance map for device length ranging length from 0.1um to 20um.

The performance map can be used to find the design with the desired performance (in this work, we randomly pick ER > 3 dB and IL < 3 dB). The related equations are in SI. Here, transmission difference is defined as the difference between the maximum transmission and minimal transmission, and it is discussed with ER and IL since it's straightforward and facilitates the analysis. The transmission difference of plasmonic mode oscillates with the device length because of periodical initial phase change caused by the various device length. (Fig. 4a). The envelope of the transmission difference with a splitting ratio of 50:50 in Fig.4b is dominated by the modulation depth and propagating loss. Both increase with device length. The envelop rises in the short length region because modulation depth increases with the device length and mainly affects the shape of the envelope. Then it decreases in the long length region because the loss of the device on the active arm increases with device length faster than the modulation depth and changes the direction of the envelope. If the propagating loss of the device on the active arm is negligible, ER reaches the largest value with a splitting ratio of 50:50 and with a specific initial phase. The initial phase can be controlled by controlling particular device lengths or using asymmetric MZI arms. (SI) However, the propagating loss of the plasmonic mode device on the active arm is not negligible. And a high splitting ratio X>50 compensates for the propagating loss but also increases the IL and reduces the output transmission. The sweet spot is usually below the splitting ratio that gives the maximal ER. The maximum ER represented by using the triangle symbol in Fig.4c is 53 dB which is achieved by the splitting ratio of 91:9. The ER looks promising because the amplitude difference of passthrough light from two arms is below $10^{-5}$ when the device is operating in an OFF state (destructive interference). However, compared with the device we proposed (length of 4.7um, speed of 108GHz, 3dB ER, 2.9dB IL, and 25%ΔT), the device with 53dB ER has a longer length of 9.1um, lower speed of 71GHz, higher IL of 10dB, and lower ΔT of 10%. High ER can be achieved by compensating the device loss (including propagating loss and design loss) using a high splitting ratio (X>50) (Fig.4d). And the splitting ratio of maximum ER design increase with device length because of increasing device loss. IL changed periodically with device length because initial phase changes and varied with splitting ratio because of increasing device loss. (Fig.4e)

For applications with limited phase shift but requires high ER, high IL is expected as a trade-off. (SI). The same concept also applies to the minimal IL device design. Based on the desired performance requirement and the nonnegligible propagating loss of the device, a low splitting ratio of 41:59 is used for the design. The modulation depth is reduced, but transmitted power is enhanced. This design perfectly balances the trade-off between the ER and IL and meets the requirements. The ER&IL map in Fig. 4f shows more design options with different device length but under the same operation voltage. The designs are distributed in a certain order. And the border marked using red solid line is mainly determined by the maximum phase change. Here we have a phase shift of about 0.33 pi. If the phase change increases, the border will move towards the shown direction.

# 3 Conclusion

In this paper, we explored material properties and the design of the modulator to reach an optimized point in terms of trade-offs between ER and IL within the constraints of small phase shifts. A 4.7μm-long ITO-plasmon-based asymmetric MZI modulator with a phase shift of 0.33π, ER of 3dB, IL of 2.9dB, speed of 108GHz under the bias of $\pm 3.5V$. The device length is carefully chosen considering high transmission difference, high ER, and low IL. Asymmetric MZI modulator (41: 59) gives rise to 0.4dB lower IL compared to the symmetric MZI modulator. Trading off modulation depth, an increased oxide thickness can reduce the device's capacitance, thus increasing the RC-limited device bandwidth. Additionally, we provide a design protocol to achieve tailored metrics such as modulation depth, speed, and losses that extend the active material of choice to engineer modulators with specific functionality-based performance [33].

# 100 GHz Micrometer-compact broadband Monolithic ITO Mach–Zehnder Interferometer Modulator enabling 3500 times higher Packing Density

**Supplementary Information**

Yaliang Gui[1], Behrouz Movahhed Nouri[1], Mario Miscuglio[1], Rubab Amin[1], Hao Wang[1], Jacob B. Khurgin[2], Hamed Dalir[1], and Volker J. Sorger[1]

[1]George Washington University, 800 22nd Street NW, Washington, DC 20052, USA

[2]Johns Hopkins University, Baltimore MD 21208, USA

[+]Corresponding authors: hdalir@gwu.edu sorger@gwu.edu

# i. Performance analysis of arbitrary power splitter based MZM (Plasmonic mode)

An asymmetric power splitter based on MZI consists of an input Y junction with a particular splitting ratio and an output Y junction with a 50:50 splitting ratio. On the active arm, a plasmonic mode ITO modulator is used to shift the phase. We will utilize a plasmonic mode ITO modulator as an example because to its advantages, which include a compact footprint, the possibility for high speed, and a large modulation depth. However, as Kramers–Kronig relations indicate, a high modulation depth is associated with a high propagating loss.

This article also discusses the effect of spreading loss. For π shift device, high ER can be achieved by compensating device propagating loss with higher splitting power (splitting ratio > 50) or using balance device on the passive arm. Low IL can be achieved at the same time by changing the device length or by using different MZI arm lengths. By using asymmetric arms, the initial phase difference, $\Delta\varphi_{initial}$, caused by different effective mode index of two arms can be precisely controlled (fig a, b,8c). Depending on the input signal, the initial phase difference, $\Delta\varphi_{initial}$, is the same with the OFF state (0V) phase difference for applications that use either positive or negative bias, while for applications that use both positive and negative bias, OFF state phase difference under negative bias would be discussed instead of the initial phase difference. Since phase difference shift forward under negative bias and vice versa. The transmission ER and IL can be calculated by using:

$$T = \frac{(\frac{\sqrt{x}}{10}E_{in}e^{-\frac{\alpha}{2}L})^2 + (\frac{\sqrt{100-x}}{10}E_{in})^2}{2} + \frac{2 \times (\frac{\sqrt{x}}{10}E_{in}e^{-\frac{\alpha}{2}L}) \times (\frac{\sqrt{100-x}}{10}E_{in}) \times \cos(\Delta\varphi_{initial} + \Delta\varphi_{bias})}{2} \quad (1)$$

$$ER = 10 \cdot \lg\left(\frac{T_{max}}{T_{min}}\right) \quad (2)$$

$$IL = 10 \cdot \lg(T_{max}) + L_{facet} \times 2 \quad (3)$$

$$\Delta\varphi_{bias} = \pi \quad (\text{For } \pi \text{ phase shift device}) \quad (4)$$

($E_{in}$, $x$, $\alpha$, $L$, $\Delta\varphi_{initial}$, $\Delta\varphi_{bias}$ and $L_{facet}$ are the input electric field phasor, the splitting ratio on the active arm side, device absorption coefficient, device length, the initial phase difference caused by different effective mode index between device and waveguide, phase changed by bias and facet coupling loss). Here we assume there is a π phase shift ITO plasmonic device with effective mode index of 2.5, $\alpha \cdot L = 0.2$ under 0 V and $\alpha \cdot L = 0.26$ under $V_\pi$. (fig. 1e and 1f) The initial phase difference is eliminated by using different MZI arms or using balanced device. $L_{facet}$ is 0.07dB obtained from Lumerical simulation. Power with high splitting ratio (x=60) on the active arm compensate the propagating loss and is used in application requires high ER. Low splitting ratio power (x=37) in contrast decrease the insertion loss by reducing the modulation depth. The overall transmissions decrease with increasing power splitter ratio because of the increasing propagating losses. Symmetric MZI with balance device has as high ER as asymmetric MZI modulator (60:40) but has 8% weaker transmission. Device with free choice of power splitting ratio is preferred than the device with balance device considering the IL.

For small phase shift device, high ER and low IL cannot be obtained at the same time. The free choice of splitting ratio provides more options for different application. High splitting ratio for applications that requires high ER, while low splitting ratio for applications with high IL requirements. What's more, the $\Delta\varphi_{initial}$ should be carefully chosen for different applications.

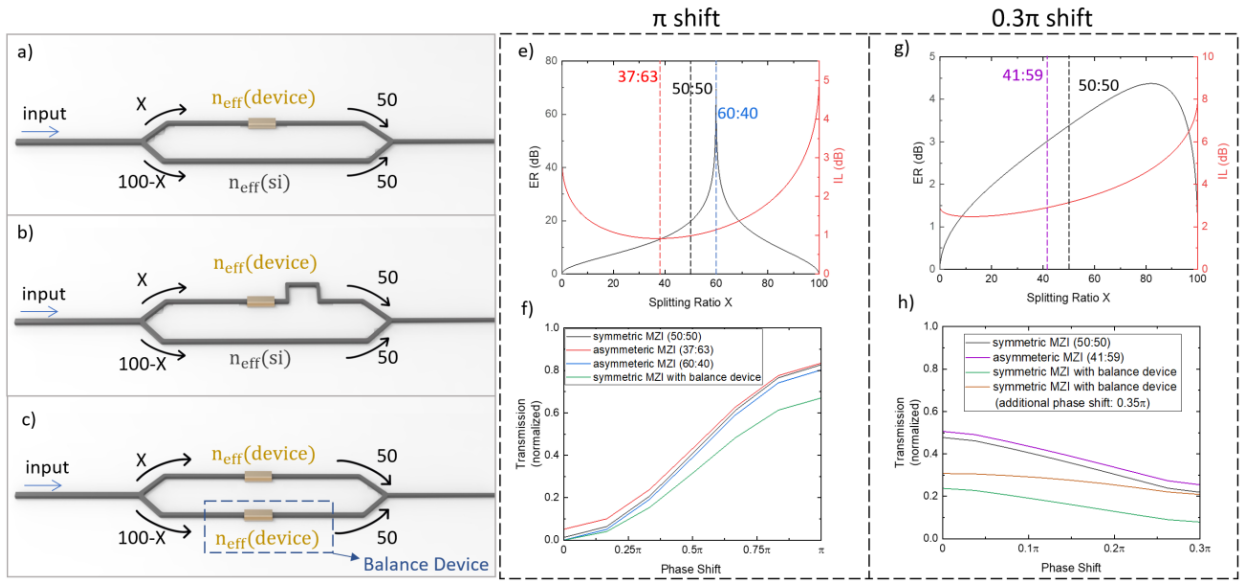

Figure 1 Performance analysis for MZI modulator with free choice power splitting ratio. a) MZI modulator with arms of the same length. b) MZI with different arm length which is used to tune the OFF-state phase shift to kπ, k=0,1,2…. c) MZI modulator with balance device on passive arm. The initial phase shift is 0. e) ER and IL of π phase shift device under different splitting ratio. (Device with power splitting ratio of 37:63 has lowest IL of 0.9dB and ER of 13dB; device with power splitting ratio of 50:50 has IL of 1dB and ER of 20dB, and device with power splitting ratio of 60:40 has IL of 1.1dB and ER of 70dB.) f) Normalized transmission of different splitting ratio (50:50, 37:63 and 60:40) and with balance device under bias with π phase shift. IL and ER of symmetric MZI modulator with balance device is 1.7dB and >70dB respectively. g) ER and IL of 0.3π phase shift device under different splitting ratio. h) Normalized transmission of different splitting ratio (50:50, 37:63 and 60:40) and with balance device under bias with 0.33π phase shift.

Here, plasmonic mode ITO-based MZI modulator 0.33π phase shift is used as an example of small phase shift device. The performances including the ER and IL is lower than previous π shift design because of weak modulation depth. However, it's useful for most of applications. The performance of symmetric MZI modulator with balance device can be optimized according to the application by using additional phase shift. Still comparing with the asymmetric MZI modulator, it's not a good choice.

## ii. Selection & Thickness of Dielectric Material

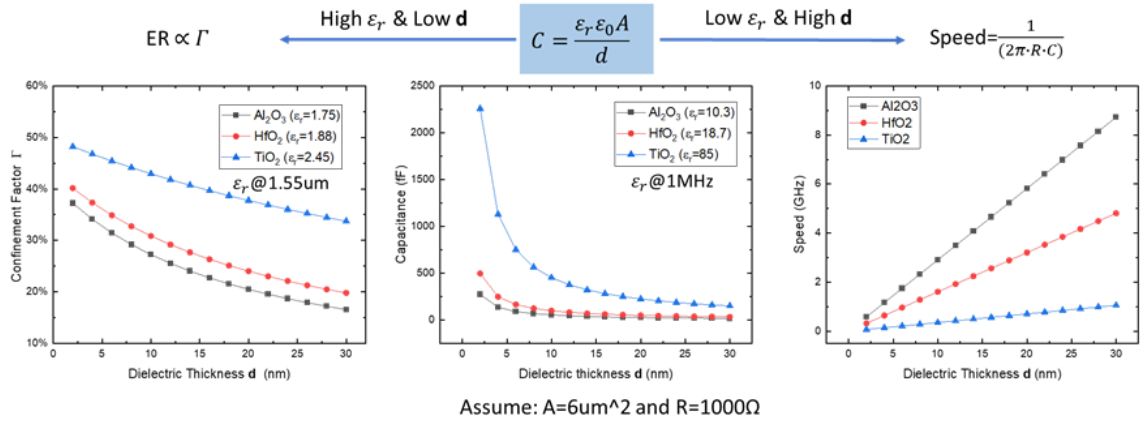

## iii. Photonic mode based on an arbitrary power splitter modulator ITO MZI

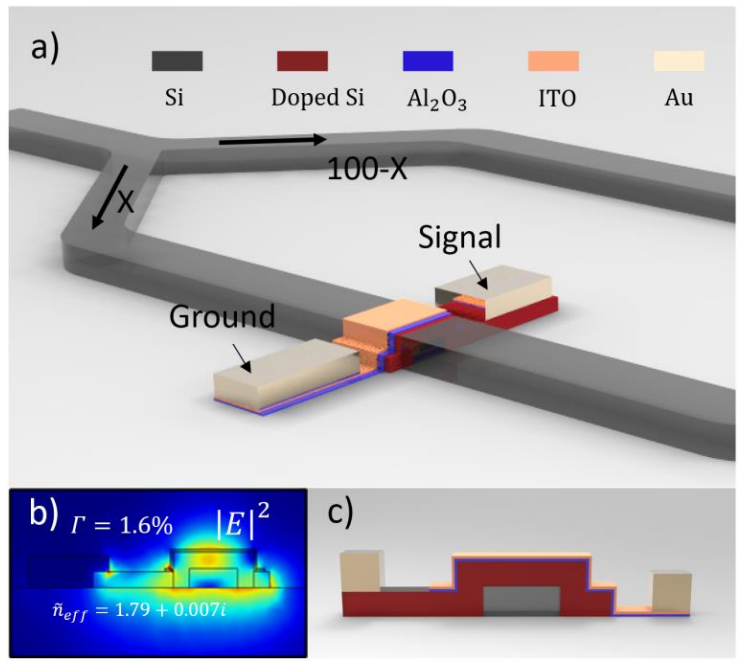

Figure 2 ITO-based Photonic mode MZI modulator on Si photonic platform. (a) Perspective view of the Mach-Zehnder structure with the active biasing contacts. The power splitting ratio for the active arm is X, and the splitting ratio for the passive arm is 100-X. (b) Electric field distribution by performing FEM at 1550nm under +3.5V in the cross-sectional structure for a z cutline along the central region of the Si waveguide (width: 500nm; height: 220nm). (c) Composition layers of ITO MZI modulator. ($T_{doped\ silicon} = 100$nm; $T_{Al_2O_3} = 10$nm; $T_{ITO} = 10$nm; $T_{Au} = 50$nm)

**Photonic mode ITO based modulator design:** The photonic mode ITO-based modulator is built on a silicon-on-insulator (SOI) platform and has an asymmetrical MZI. On the input side, a Y-junction has been carefully designed as a power splitter with an arbitrary (non-50:50) splitting ratio determined by device loss and a 50:50 Y-junction on the output side to balance the power of each arm. For the photonic mode device option, silicon is doped at the device area as the bottom contact with a doping concentration of 1020 cm$^{-3}$, resulting in a conductivity of 97087 S/m. The doping depth must not be too thin in order to maintain a low resistance and was chosen in this case to be 100nm. At 1550nm, the complex refractive index[1] of doped silicon is calculated as follows:

$$n_{doped} = n_{intrinsic} + \Delta n$$

$$k_{doped} = k_{intrinsic} + \Delta k$$

$$\Delta \alpha = (4 \cdot \Delta k \cdot \pi)/\lambda$$

$$\Delta \alpha = \Delta \alpha_e + \Delta \alpha_h = 8.88 \times 10^{-21} \times \Delta N_e^{1.167} + 5.84 \times 10^{-20} \times \Delta N_h^{1.109}$$

$$-\Delta n = \Delta n_e + \Delta n_h = 5.4 \times 10^{-22} \times \Delta N_e^{1.011} + 1.53 \times 10^{-18} \times \Delta N_h^{0.838}$$

Where, $\Delta \alpha_e, \Delta \alpha_h, \Delta n_e, \Delta n_h$, are absorption coefficient change by doped electrons and holes, respectively, and the refractive index change again arising from doped electrons and holes, number of electrons introduced through doping, respectively. $\Delta N_e \ and \ \Delta N_h$ are the carrier concentrations for both charges introduced by doping. $n_{intrinsic}$=3.48 and $k_{intrinsic} = 0$. The carrier concentration of unbiased ITO thin film is $2.07 \times 10^{20} cm^{-3}$. Under ±5V, the refractive indexes for 10nm ITO thin film are 1.7896+0.007686i and 1.6539+0.00261i. a 10nm thin Al$_2$O$_3$ film between ITO thin film and doped silicon region is used as gating oxide. The anticipated speed is 108GHz.

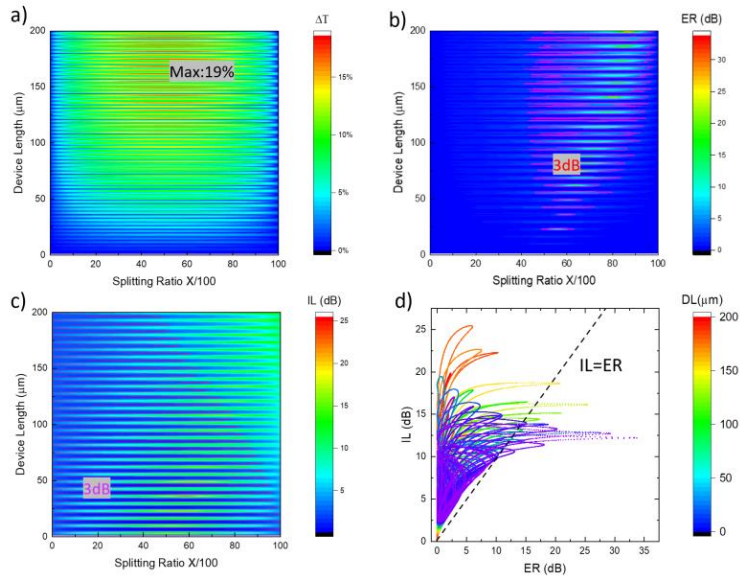

Figure 3 The performance map of asymmetric photonic mode ITO-based MZI modulator with 15nm $Al_2O_3$ under ON (-5V) and OFF states (+3.5V). a) Transmission difference varies with splitting ratio X and device length 0 $\mu m$ to the maximum transmission difference of the photonic mode device is 19% because of weak modulation depth. Photonic mode designs with high transmission differences locate in long-length regions because of weak modulation depth and low device loss. b) Extinction ratio of photonic device varies with splitting ratio for different lengths. c) Insertion loss of plasmonic mode and photonic mode varies with splitting ratio for different lengths. d) Performance map for device length ranging length from 0.1um to 20um.